\documentclass[preprint]{aastex}
\usepackage{epsf}
\usepackage{emulateapj5}
\usepackage{onecolfloat}
\usepackage{apjfonts}
\usepackage{amsmath}

\def\gtsim {\lower .1ex\hbox{\rlap{\raise .6ex\hbox{\hskip .3ex
        {\ifmmode{\scriptscriptstyle >}\else
                {$\scriptscriptstyle >$}\fi}}}
        \kern -.4ex{\ifmmode{\scriptscriptstyle \sim}\else
                {$\scriptscriptstyle\sim$}\fi}}}
\newcommand{\beq}{\begin{equation}}
\newcommand{\eeq}{\end{equation}}

\def\Mo{{\rm M_\odot}}

\def\kpc{\ {\rm kpc}}

\def\LCDM{$\Lambda$CDM}

\begin{document}
\submitted{The Astrophysical Journal, accepted}
\vspace{1mm}
\slugcomment{{\em The Astrophysical Journal, accepted}}

\shortauthors{KAZANTZIDIS ET AL.}
\twocolumn[
\lefthead{SHAPES OF DARK MATTER HALOS}
\righthead{KAZANTZIDIS ET AL.}

%----------------------------------------------------------------
\title{The effect of gas cooling on the shapes of dark matter halos}
%----------------------------------------------------------------

\author{Stelios Kazantzidis,\altaffilmark{1,2} 
  Andrey V. Kravtsov,\altaffilmark{2} 
  Andrew R. Zentner,\altaffilmark{2}
  Brandon Allgood,\altaffilmark{3}\\
  Daisuke Nagai,\altaffilmark{2} 
  and Ben Moore\altaffilmark{1}
  }
\vspace{2mm}

\begin{abstract}
  We analyze the effect of dissipation on the shapes of dark matter (DM) 
  halos using high-resolution cosmological gasdynamics
  simulations of clusters and galaxies in the {\LCDM} cosmology.
  We find that halos formed in simulations with gas cooling are
  significantly more spherical than corresponding halos formed in
  adiabatic simulations. Gas cooling results in an average increase
  of the principle axis ratios of halos by $\sim 0.2-0.4$ in the inner
  regions. The systematic difference decreases slowly with radius but
  persists almost to the virial radius. We argue that the differences
  in simulations with and without cooling arise both during periods of
  quiescent evolution, when gas cools and condenses toward the center,
  and during major mergers. We perform a series of high-resolution
  $N$-body simulations to study the shapes of remnants in major
  mergers of DM halos and halos with embedded stellar disks. In the DM
  halo-only mergers, the shape of the remnants depends only on the
  orbital angular momentum of the encounter and not on the internal
  structure of the halos. However, significant shape changes in the DM
  distribution may result if stellar disks are included. In this case
  the shape of the DM halos is correlated with the morphology of the
  stellar remnants.  

\end{abstract}

 ]

\altaffiltext{1}{Institute for Theoretical Physics, University of Z\"urich, 
Winterthurerstrasse 190, CH-8057 Z\"urich, Switzerland; stelios@physik.unizh.ch.}
\altaffiltext{2}{Department of Astronomy and Astrophysics,
       Kavli Institute for Cosmological Physics, 5640 South Ellis Avenue,
       University of Chicago, Chicago, IL 60637.}
\altaffiltext{3}{Department of Physics, University of California at Santa Cruz,
	1156 High Street, Santa Cruz, CA 95064.}

%--------------------------------------------------
\section{INTRODUCTION}
\label{section:introduction}
%-------------------------------------------------

Triaxial dark matter (DM) halos are a generic prediction of the
hierarchical, cold dark matter (CDM) model of structure formation.
Therefore, observational probes of halo shapes are fundamental 
tests of this model. Recently, such comparisons have 
received attention as the accuracy and number of observational
probes of halo shapes increase for the Milky Way (MW) 
\citep[see][for a recent review]{merrifield03}, other galaxies
\citep[e.g.,][]{buote_canizares98,buote_etal02,hoekstra_etal04}, and
galaxy clusters \citep[e.g.,][]{kolokotronis_etal01}.

\citet{dubinski_carlberg91} first pointed out that 
observed elliptical galaxies are systematically more spherical than 
simulated CDM halos. It was recently argued that the coherence of the tidal
stream of the Sagittarius dwarf spheroidal galaxy indicates that the 
inner halo of the MW is nearly spherical, with a minor-to-major axis 
ratio $c/a \gtrsim 0.8$ \citep[][but see \citeauthor{mayer_etal02} \citeyear{mayer_etal02}, 
\citeauthor{helmi03} \citeyear{helmi03}, and \citeauthor{martinez_delgado_etal04}
\citeyear{martinez_delgado_etal04}]{ibata_etal01,majewski_etal03}.
This differs considerably from an average predicted ratio of
$\langle c/a \rangle \sim 0.6-0.7$ for MW-sized halos formed in {\em
dissipationless} cosmological simulations 
\citep{barnes_efstathiou87,frenk_etal88,dubinski_carlberg91,warren_etal92,
cole_lacey96,bullock02,jing_suto02}.  Although the axis ratios of simulated halos
exhibit a broad distribution, with a Gaussian rms dispersion of
$\sim 0.1$, we would have to accept the MW halo as an outlier.

Utilizing halo shapes as a test of the CDM paradigm requires improving
both observations and theoretical predictions.  Most of the numerical
work published to date does not account for the effects of gas cooling
on halo shapes.  Notable exceptions are the studies by \citet{katz_gunn91},
\citet{evrard_etal94}, and \citet{tissera_dominguez_tenreiro98}, who
find that halos in dissipational simulations are systematically
more spherical than corresponding halos in dissipationless runs.
However, the halos in these studies were resolved with only a few
hundred to a few thousand particles and the shapes of the inner
regions of halos could not be studied reliably. In addition, these
studies did not address the processes by which cooling affects halo shapes.
Recently, \citet{floor_etal03} used gasdynamics simulations of galaxy 
clusters with and without radiative cooling to investigate the evolution of 
eccentricity. Their study showed very slow eccentricity evolution in simulated 
clusters. However, they focused on the eccentricity at large radii where,
as we demonstrate, the effect of cooling is small.

\citet{dubinski94} studied the effects of dissipation on halo
shapes numerically by adiabatically growing central mass concentrations 
in an initially triaxial $N$-body halo. This study showed that the growth of a 
central condensation results in a considerably rounder halo shape.
These experiments, although suggestive, were not generally representative of the 
cosmological framework of hierarchical halo assembly via multiple mergers.
 
In this Letter we revisit the effects of dissipation on halo
shapes, combining self-consistent cosmological simulations with
a resolution up to $2$ orders of magnitude higher than in previous
studies, with high-resolution controlled $N$-body simulations of binary
mergers of disk galaxies and DM halos whose structure is based on the 
currently favored galaxy formation paradigm.  

%--------------------------------------------------
\section{NUMERICAL SIMULATIONS}
\label{section:num_sim}
%--------------------------------------------------

We analyze high-resolution cosmological simulations of eight group- and
cluster-size systems and one galaxy-size system in a flat {\LCDM}
model: $\Omega_{\rm m}=1-\Omega_{\Lambda}=0.3$, $\Omega_{\rm
  b}=0.043$, $h=0.7$ and $\sigma_8=0.9$.  The simulations were 
performed with the Adaptive Refinement Tree 
$N$-body$+$gasdynamics code \citep{kravtsov99, kravtsov_etal02}, an
Eulerian code that uses adaptive refinement in space and time and
(nonadaptive) refinement in mass to achieve the high dynamic range
required to resolve cores of halos formed in cosmological simulations.

The cluster simulations have a peak resolution of $\approx 2.44\,h^{-1}$
kpc and DM particle mass of $2.7\times 10^{8}\,h^{-1}{\rm\ 
  M_{\odot}}$ with only a region of $\sim 10\,h^{-1}\ \rm Mpc$ around
each cluster adaptively refined.  We analyze each cluster at an epoch
near $z=0$, when it appears most relaxed. This minimizes the
noise introduced by substructure, which may cause transient changes in
the estimates of axis ratios at particular radii.  For one cluster, we
also analyze its progenitors at $z=1$ and $z=2$.  The virial
masses\footnote{We define the virial radius as the radius enclosing an
  average density of $180$ times the mean density of the universe at the
  epoch of analysis.}  of the clusters range from $\approx 10^{13}$ to 
$3\times 10^{14}\,h^{-1}{\ \rm M_{\odot}}$.  The galaxy formation simulation follows the early
($z \geq 4$) evolution of a galaxy that becomes an MW-size object at
$z=0$ in a periodic box of $6\,h^{-1}\ \rm Mpc$.  At $z=4$, the galaxy
already contains a large fraction of its final mass: $\approx 2 \times
10^{11}\,h^{-1}\ \rm M_{\odot}$.  The DM particle mass is 
$9.18\times 10^5\,h^{-1}\ \rm M_{\odot}$ and the peak resolution of
the simulation is $183\,h^{-1}$~comoving pc.  This simulation is
presented in \citet{kravtsov03}, where more details can be found.

For each object, we analyze two sets of simulations started from the
same initial conditions but including different physical processes.
The first set of simulations follows the dynamics of gas
``adiabatically,'' i.e., without radiative cooling.  For the cluster
halos we also analyze purely dissipationless simulations without gas 
the results of which agree well with the adiabatic gasdynamics runs.  
The second set of simulations includes star formation, 
metal enrichment and thermal supernovae feedback,
metallicity- and density-dependent cooling, and UV heating due to a
cosmological ionizing background. We use these cosmological
simulations to study the effect of dissipation on the shapes of 
halos in the next section. 

To explore the mechanisms responsible for the effect of cooling on halo
shapes, we performed controlled $N$-body merger simulations of equal-mass 
pure DM halos and multicomponent galaxies varying only
the internal mass distributions (Kazantzidis et al., in preparation).
The DM halos followed the \citet[][hereafter NFW]{navarro_etal96}
density profile and were initially spherical with isotropic velocity
dispersion tensors \citep{kazantzidis_etal04}.  First, we simulated 
several mergers of systems on parabolic orbits with pericenters 
that were 20\% of the halo virial radii, typical of cosmological mergers
\citep{khochfar_burkert03}.  In one set of experiments, we merged DM halos with
embedded stellar disks \citep{hernquist93,mo_etal98}. In these experiments
we used randomly inclined and coplanar disk configurations and adopted halo parameters 
from the MW model A1 of \cite{klypin_etal02}. Specifically, each halo had a
virial mass of $M_{\rm vir}=7 \times10^{11}\,h^{-1} \Mo$, a
concentration parameter of $c=12$, and a dimensionless spin parameter
of $\lambda=0.031$. The halos were adiabatically contracted to
respond to the growth of the stellar disk.  The mass and thickness of the 
stellar disk were $M_{\rm d}=0.04 M_{\rm vir}$ and $z_{0}=0.1 R_{\rm d}$, respectively, 
and $R_{\rm d}=2.7\,h^{-1}\kpc$ was the disk scale length.
In another experiment, we studied the merging
of identical NFW halos with no disk component. In a third experiment, we
studied the merging of ``contracted halos'' with steepened inner density 
profiles ($\rho \sim r^{-1.7}$). These profiles were set to match the combined 
initial spherical density profile of the adiabatically contracted halos and 
stellar disks in the halo$+$disk mergers. Last, we repeated the 
halo$+$disk and pure NFW halo mergers for radial orbits 
to study the effect of the impact parameter on the structure of the remnant \citep{moore_etal03}.
The initial separation of the halo centers was twice their virial radii. 
Their initial relative velocity was determined from the corresponding Keplerian orbit of 
two point masses for the parabolic mergers, while for the radial mergers it 
was set to the velocity of the circular orbit about their common center of mass.

The merger simulations were performed with the PKDGRAV \citep{stadel01}, 
multistepping, parallel $N$-body tree code. We used $N=2 \times 10^5$ 
DM particles and $N=2 \times10^4$ collisionless stellar particles and 
employed a gravitational softening of $0.7$ and $0.3\,h^{-1}\kpc$, respectively.
Our results did not change when we used a factor of $10$ more particles 
and half the softening indicating the achievement of numerical convergence.
\begin{figure}[t]
\centerline{\epsfxsize=3.5in \epsffile{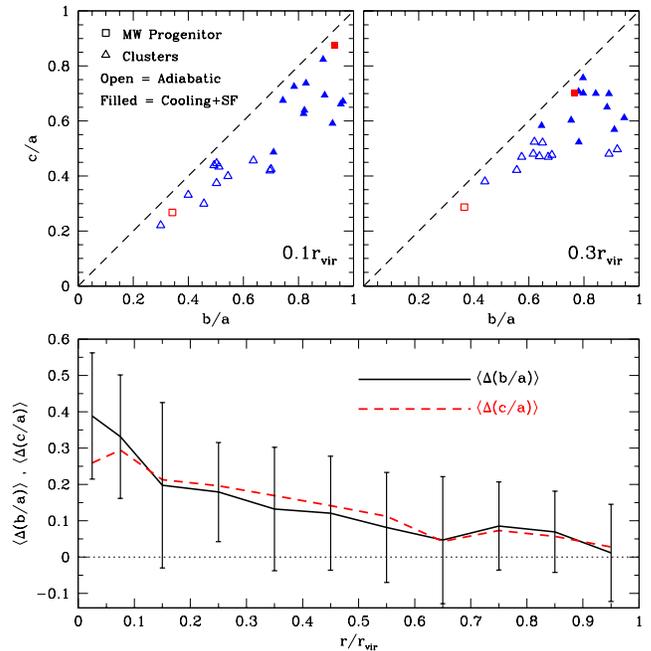}}
\caption{{\it Top panels:} Axis ratios $c/a$ vs. $b/a$ at 10\% and 
  30\% of the virial radius [using bins of $\Delta(r/r_{\rm vir}) = 0.1$]
  for halos in cosmological simulations.  Open symbols show adiabatic
  simulations and filled symbols correspond to simulations
  with gas cooling and star formation. The galaxy simulation is
  shown by a square and the clusters by triangles. 
  {\it Bottom panel:} Average difference between
  axis ratios in cooling and adiabatic simulations as a function of radius.
  The error bars show the $1\sigma$ scatter about the mean value of 
   $\Delta(b/a)$ in each bin. The scatter in $\Delta(c/a)$ is similar.
\label{fig1}}
\end{figure}
%
%-------------------------------
\section{RESULTS}
\label{section:results}
%--------------------------------

For each object, we calculate principle axis ratios $s = b/a$ and $q =
c/a$ ($a>b>c$), from the eigenvalues of a modified inertia tensor
\citep[e.g.,][]{dubinski_carlberg91}: $I_{ij} = \sum_{\alpha}
x_{i}^{\alpha}x_{j}^{\alpha}/r^{2}_{\alpha}$, where $x_{i}^{\alpha}$
is the $i$ coordinate of the $\alpha$th particle, $r^2_{\alpha} =
(y^{\alpha}_{1})^2 + (y^{\alpha}_{2}/s)^2 + (y^{\alpha}_{3}/q)^2$, and
$y^{\alpha}_{i}$ are coordinates with respect to the principle
axes. We use an iterative algorithm starting with a spherical
configuration ($a=b=c$) and use the results of the previous
iteration to define the principle axes of the next iteration until the
results converge to a fractional difference of $10^{-3}$.  We compute
axis ratios ``differentially'' using only particles within finite bins
of $r$, and label each bin by the mean value of $r$ of
the particles in the bin.  We use differential measurements because
the axis ratios calculated at different values of $r$ are almost
independent.  We find that in the standard practice of measuring
shapes cumulatively and weighting each particle contribution to
$I_{ij}$ by $r^{-2}$, axis ratios at large $r$ are quite sensitive
to the distribution of particles in the inner regions of the halo. The
differences between the differential and cumulative estimates of axis
ratios can be as large as a few tenths at large radii. However, we
present shape measurements in differential radial bins. Within each
bin, different definitions of $I_{ij}$ result in very small differences. 

\begin{figure}[t]
\centerline{\epsfxsize=3.5in \epsffile{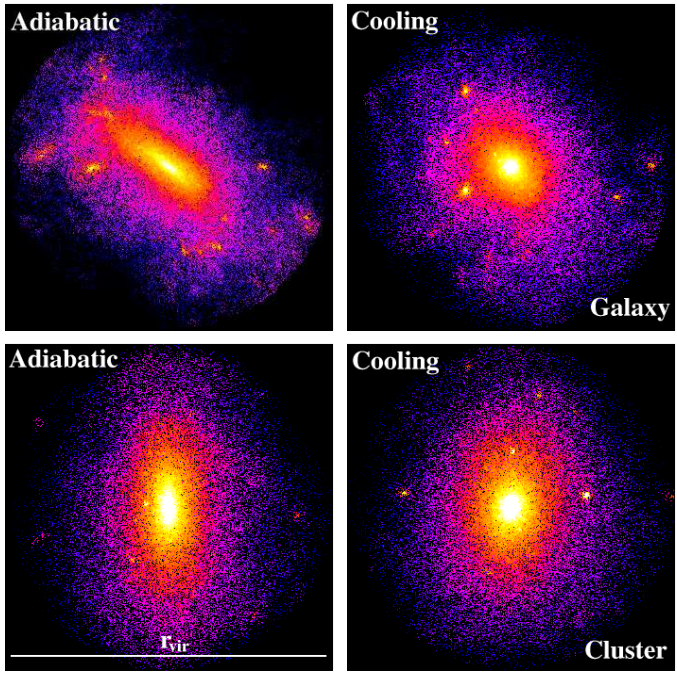}}
\caption{Density maps for the galaxy-size ({\it top panels})
  and a cluster-size ({\it bottom panels}) halo projected onto the
  plane of intermediate and major axis. The particles are color-coded
  on a logarithmic scale with brighter colors in higher density
  regions. The local density is calculated using an SPH smoothing kernel of 
  $64$ particles to suppress small substructure and to better show the shape of 
  the diffuse DM in the main halo. Halos in the cooling runs {\it(right panels)} 
  are significantly rounder than their adiabatic counterparts {\it (left panels)}.
\label{fig2}}
\end{figure}

Figure~\ref{fig1} shows the principle axis ratios of the {\it dark matter} 
at 10\% and 30\% of the virial radius for all cosmological simulations.  
The purely dissipationless simulations without gas 
produce results similar to the adiabatic runs and are omitted for clarity.  
Note that the halos in adiabatic simulations tend to be slightly 
more elongated closer to their centers.  Interestingly, there is a 
correlation of $c/a$ and $b/a$.  The effect of cooling changes axis 
ratios roughly along the correlation in the $c/a$-$b/a$ plane by up 
to $\sim 0.4$ at $0.1r_{\rm vir}$ and by $\sim 0.2-0.3$ at $0.3r_{\rm vir}$.  
The significance of the change in shape can be seen in Figure~\ref{fig2},
where we show density maps of the DM distribution
in the simulated galaxy ({\it top panels}) and a cluster ({\it bottom panels}) 
at $z=4$ and $z=0.4$, respectively.  
The difference between the adiabatic and cooling simulations is 
visually striking, with the DM halos significantly rounder in the 
cooling runs.

\begin{figure}[t]
\centerline{\epsfxsize=3.2in \epsffile{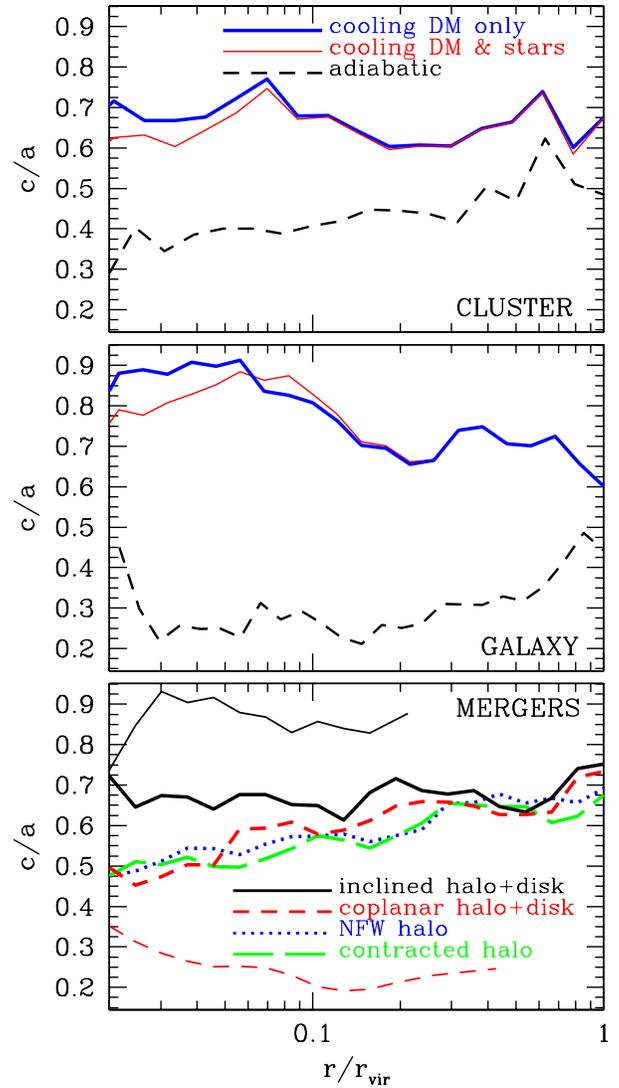}}
\caption{ {\it Top and middle panels:}  Minor-to-major axis ratio, $c/a$,
  as a function of radius for a cluster-size ({\it top panel}) and the
  galaxy-size ({\it middle panel}) halo. The thick solid lines
  correspond to the DM, while the thin solid lines show the
  combined $c/a$ of DM and stars in simulations with cooling. The
  dashed lines show the profile of the DM in the adiabatic
  simulations.  {\it Bottom panel:} The $c/a$ profiles of merger remnants.
  We show remnants from four mergers: inclined
  halo$+$disk ({\it solid line}), coplanar halo$+$disk ({\it short-dashed line}), 
  NFW halo ({\it dotted line}), and contracted halo ({\it long-dashed line}).  
  Thick lines show $c/a$ for DM only, and thin lines show $c/a$ for stars in the cases with
  initial stellar disks in the inclined ({\it solid line}) and coplanar ({\it short-dashed line}) 
  mergers.  We show the profiles at a time $\sim 8$ crossing
  times of the remnant.
\label{fig3}}
\end{figure}

The radial dependence of the effect can be seen in the bottom panel of
Figure~\ref{fig1}, which shows the average difference between axis
ratios in the cooling and adiabatic runs 
[$\Delta(b/a) \equiv (b/a)_{\rm cool} - (b/a)_{\rm adiab}$] 
as a function of radius.
Although the effect decreases with radius, the axis ratios in
simulations with cooling are systematically larger than in the
adiabatic and dissipationless cases, even at $r \approx 0.8 r_{\rm vir}$.
Figure~\ref{fig3} shows individual radial profiles of the minor-to-major 
axis ratio $c/a$ in the galaxy and cluster simulations, as well as in
representative controlled merger experiments. It is interesting to
compare cluster and galaxy halos because their baryon distributions 
have different morphologies. In the galaxy, 
most ($\sim 90\%$) of the baryons lie in a highly flattened gaseous disk, 
while in the cluster simulations most of the baryons are in stars 
in a massive, central elliptical galaxy.
For both the galaxy and the cluster systems the effect of cooling 
is only weakly dependent on radius and is significant even at $r_{\rm vir}$.
Note that axis ratios are not constant as a function of radius and
that changes are not monotonic.  Different regions of an object
may be flattened to different degrees. This indicates that 
observational estimates of halo shapes in the disk region
\citep[e.g., by disk flaring, ][]{olling_merrifield00} may be somewhat
different from results at larger distances.  Note 
that despite the significant flattening of the baryons in the central
$0.1r_{\rm vir}$ of the galaxy simulation, the DM
distribution around the disk is almost spherical.

It is interesting to ask if the ellipsoid of the DM halo is aligned
with that of the baryons. For clusters, in which most of the
baryons in the center are in stars, the major axes of the stellar and DM 
distribution are approximately aligned at all radii. For example, 
the major axis of the central cluster galaxy is well aligned with the 
inner DM halo.  However, we find that in clusters that contain 
massive substructures in their outer regions, the direction of 
the major axis often changes dramatically at $r\gtrsim 0.5r_{\rm vir}$.  
In the galaxy simulation, the minor axis of the DM distribution in the 
vicinity of the disk is aligned with the minor axis of the
disk. However, the flattening of the DM distribution is small
(Figure~\ref{fig3}). Interestingly, at $r\gtrsim 0.2r_{\rm vir}$, the
direction of the major axis of the DM halo changes and is nearly 
perpendicular to the disk.

We examined the evolution of the merger remnants and found that their 
shapes evolve in their outer regions for $\sim 8$ crossing
times or $\sim 14-18$~Gyr.  This indicates that the shapes of the 
outer regions of the cosmological halos are evolving at all epochs.  
The bottom panel of Figure~\ref{fig3} shows the axis ratios of
the remnants in controlled merger experiments after $8$ crossing times, 
when the evolution has ceased. The shapes of the DM halo merger 
remnants vary, depending strongly on the 
presence of a disk component and the relative inclination of disks 
prior to the merger.  Mergers of halos with different central density 
profiles produce remnants with very similar axis ratios.  When a 
disk component is present, the shape of the remnant DM halo depends 
sensitively on the initial relative inclination of the disks.  
Inclined disk mergers lead to a very spherical stellar remnant and 
a correspondingly more spherical DM halo, compared to the halo-only cases. 
Coplanar disk mergers lead to a very disklike stellar component 
(small $c/a$) and a DM halo that is nearly as flattened as in the 
halo-only mergers.  We discuss the implications of these results 
in the next section.

%------------------------------------
\section{DISCUSSION and CONCLUSIONS}
\label{section:discussion_conclusions}
%------------------------------------

We show that halos in cosmological simulations with cooling are 
considerably more spherical than those in dissipationless simulations.  
The difference decreases with increasing radius 
but can be significant even at the virial radius.  This is somewhat
surprising because cooling affects the mass distribution appreciably
only in the inner $\sim 10\%$ of the virial radius. 
The condensation of baryons due to cooling leads to a more
concentrated distribution of DM, as it responds to the
increasing gravitational field of baryons in the center
\citep{blumenthal_etal86}. Thus, dissipation results in a
significantly more centrally concentrated mass distribution and
a deeper gravitational potential. \citet{dubinski94} showed that this
leads to the evolution of the halo toward a more spherical shape in a 
few crossing times, arguing that as the central
condensation grows, the overall potential becomes rounder. 
This shifts the boundaries between orbital families markedly, decreasing the
fraction of regular box orbits that serve as the backbone of a triaxial 
mass distribution \citep{gerhard_binney85,udry_martinet94,barnes_hernquist96,
merritt_quinlan98,valluri_merritt98}.

In hierarchical models of structure formation, halos grow via a sequence of violent 
mergers and periods of slow accretion. Although cooling can gradually make a
halo more spherical, subsequent mergers can produce highly elongated
remnants \citep[e.g.,][]{moore_etal03}, erasing the effect of
dissipation discussed above. If no significant cooling occurs after
the last major merger,\footnote{For example, if the merger occurs after
  most of the gas is converted to stars or the cooling time in the
  merger remnant is long.} the triaxiality of the halo will be largely
determined by the merger. Hence it is important to consider how
cooling affects the shapes of merger remnants. To this end, we
analyze a suite of controlled merger simulations of pure DM halos and
halos with embedded disks.

Cooling can directly affect the shape of stellar remnants during
mergers \citep{barnes_hernquist96}.  However, a large amount of
cooling gas may be needed for this to significantly affect the shapes
of DM halos. Indeed, we compare remnant shapes in mergers of
disk$+$halo systems in which disks contain both stars and a modest
amount of gas (10\% of the total disk mass), with and without cooling.
This comparison shows that the effect of dissipation on the shapes of
dark halos during the mergers of these stellar-dominated disks is
negligible.

On the other hand, our merger simulations demonstrate that the shape
of the remnant DM halo is affected by the presence of cold
stellar disks. Figure~\ref{fig3} shows that the shape of the DM
distribution in the remnant is always correlated with the shape of the
stellar distribution.  Inclined disk mergers produce a nearly
spherical stellar remnant, while coplanar disk mergers yield a very 
oblate stellar remnant with $c/a\sim 0.3$.  The
corresponding DM distribution is also more spherical in the inclined
disk merger.  The shape of the stellar remnant is thus quite sensitive
to the mutual orientation of the merging disks. It is likely to be
also sensitive to their internal properties. Indeed, in an inclined halo$+$disk merger 
with the same orbital parameters as before, but with a factor of $3$ thicker 
stellar disks, the stellar remnant is less spherical and the halo axis 
ratios are lower by $\sim 0.15$ in the inner $r\lesssim 0.1r_{\rm vir}$.
If a spherical remnant is formed, the potential in the central region, where stars dominate
gravitationally, will also be spherical. This likely destroys the regular
box orbits of DM particles and drives the DM distribution to
a more spherical configuration. This effect is not confined to the 
vicinity of the stellar remnant. The DM distribution may be affected
at larger radii because DM particles in CDM halos are, on average, on
eccentric orbits with a median apocenter-to-pericenter ratio of 
$r_{\rm apo}/r_{\rm per} = 6/1$ \citep{ghigna_etal98}.

Our results indicate that the effect of cooling during quiescent
evolution and its indirect effect during late-time disk mergers 
are both important in determining the shape of the final
mass distribution. These effects lead to halos that are
systematically more spherical than those measured in
dissipationless CDM simulations. This has several important
implications for comparisons with, and interpretations of, observations.
The mean axis ratios of MW-size halos may be expected to be $\langle
c/a\rangle\sim 0.7-0.8$ in their inner regions compared to 
$\langle c/a\rangle\sim 0.6-0.7$ estimated from dissipationless $N$-body 
simulations.  This can explain the observed coherence of the tidal streams of the
Sagittarius dwarf spheroidal galaxy without requiring the MW halo to be an 
outlier in the distribution of halo shapes.  We may also expect any intrinsic
correlations between the shapes of neighboring galaxies to be mitigated.  
The effect of triaxiality on the average lensing cross section 
of clusters may not be as strong as previously
thought \citep{dalal_etal03,oguri_keeton04}. More spherical inner
halos may influence the dynamics of bars and stars in galactic disks
\citep[e.g.,][]{el-zant_shlosman02,gadotti_desouza03} and may limit the mass of nuclear
supermassive black holes by inhibiting their fueling \citep{merritt_quinlan98}.

The effect of cooling is likely to be mass dependent because the
efficiency of cooling and star formation, manifested in the average
mass-to-light ratio $M/L$, is a fairly strong non-monotonic function
of mass \citep[e.g.,][]{vandenbosch_etal03}. Current cosmological
simulations that include gasdynamics suffer from the ``overcooling
problem'': the fraction of baryons in stars and cold gas is at least a
factor of $2$ higher than observed for the systems of the mass range
that we consider \citep{lin_etal03}. The effect on the axis ratios is
therefore likely overestimated and substantial improvements in
cosmological simulations are required to predict the shape
distribution of CDM halos accurately.  Nevertheless, the effect that we
find is so strong that even with significantly reduced fractions
of cold gas, we expect the change in the axis ratios in the inner
regions of halos to be $\gtrsim 0.1$, comparable to or larger than the
dispersion in the distribution of measured axis ratios in
dissipationless $N$-body simulations.

\acknowledgments

We thank Victor Debattista, John Dubinski, Lucio Mayer, and Monica
Valluri for useful discussions. B. M. would like to thank John
Magorrian for valuable comments on shape and orbital structure.  S. K.
is grateful to the Kavli Institute for Cosmological Physics at the
University of Chicago for the hospitality during his visit.  This work
is supported by the NSF and NASA under
grants AST 02-06216, AST 02-39759, NAG5-13274, and NAG5-7986 and
by the Kavli Institute for Cosmological Physics at the University of Chicago
and the Institute for Theoretical Physics at the University of
Z\"urich. D. N. is supported by the NASA Graduate Student Researchers
Program. The cosmological simulations used in this study were performed on 
the IBM RS/6000 SP4 system at the National Center for Supercomputing Applications.
The merger simulations were carried out on the Intel cluster at the
Cineca Supercomputing Center in Bologna, Italy.

\bibliography{shapes}

\end{document}